\documentclass[a4paper,fleqn,usenatbib]{mnras}

\usepackage{newtxtext,newtxmath}

\usepackage[T1]{fontenc}
\usepackage{ae,aecompl}

\usepackage{graphicx}	
\usepackage{amsmath}	
\usepackage{amssymb}	
\usepackage{siunitx}	
\usepackage[table]{xcolor} 
\usepackage{multicol, caption}
\usepackage{array}

\newcolumntype{C}[1]{>{\centering\arraybackslash}p{#1}}

\title[Environments of dwarfs with AGN]{Environments of dwarf galaxies with optical AGN characteristics}

\author[M. T. Kristensen et al.]{
Mikkel T. Kristensen,$^{1}$\thanks{E-mail: m.t.kristensen-2018@hull.ac.uk}
Kevin Pimbblet,$^{1}$
and Samantha Penny$^{2}$
\\
$^{1}$E. A. Milne Centre for Astrophysics, University of Hull, Cottingham Road, Kingston-upon-Hull HU6 7RX, UK\\
$^{2}$Institute of Cosmology and Gravitation, University of Portsmouth, Dennis Sciama Building, Burnaby Road, Portsmouth PO1 3FX, UK
}

\date{Accepted XXX. Received YYY; in original form ZZZ}

\pubyear{2020}

\begin{document}
\label{firstpage}
\pagerange{\pageref{firstpage}--\pageref{lastpage}}
\maketitle

\begin{abstract}
This study aims to explore the relation between dwarf galaxies ($M_* \leq 5\times10^9 M_\odot$) with AGNs and their environment by comparing neighbourhood parameters of AGN and non-AGN  samples. Using the NASA-Sloan Atlas, both the local environment and the immediate environment of dwarf galaxies with $z \leq 0.055$ are analysed. Of the 145,155 galaxies in the catalogue, 62,258 of them are classified as dwarf galaxies, and by employing two AGN selection methods based on emission line fluxes (BPT and WHAN), 4,476 are found to have AGN characteristics in their optical spectra. Regardless of selection method, this study finds no discernible differences in environment between AGN and non-AGN host dwarf galaxies and these results indicate that environment is not an important factor in triggering AGN activity in dwarf galaxies. This is in line with existing literature on environments of regular galaxies with AGNs and suggests universality in terms of reaction to environment across the mass regime. The biases of AGN selection in low-mass galaxies, and the biases of different measures of environment are also considered. It is found that there are several mass-trends in emission line ratios and that the SDSS fiber covers galaxies non-uniformly with redshift. These biases should be accounted for in future work by possibly including other wavelength regimes or mass-weighting of emission line ratios. Lastly, a discussion of the environment estimation methods is included since they may not gauge the desired properties due to factors such as time delay or using loosely constrained proxy parameters.
\end{abstract}




\section{Introduction}
\label{sec:introduction}
Galaxies are dynamical objects that evolve and mature over time. Internal processes such as star formation, supernovae, and nuclear activity and external ones such as galaxy interactions \citep{Moore1996}, ram-pressure stripping \citep{Gunn1972}, and intergalactic medium accretion can change the composition and structure of galaxies and decide their futures. Many of these processes are strongly correlated with stellar mass or environment \citep{Kauffmann2003, Miller2003, Baldry2006, Peng2010, Peng2012}.

Multiple processes can affect galaxies simultaneously. Dwarf galaxies can be used to isolate a single evolutionary process due to their low masses and relatively low frequency of mergers. These properties potentially give a single process a huge impact on the evolution of them. For example, field dwarfs are very much shaped only by internal processes while environmental effects dominate low-mass galaxies in clusters and groups \citep{Haines2007, Peng2010}. 

Observing and analysing dwarf galaxies is observationally expensive and time consuming since their low surface brightness require long exposures. For example, in a survey similar to the Sloan Digital Sky Survey \citep[SDSS][]{York2000}, a galaxy such as the Large Magellanic Cloud (LMC) will only be observable out the $z \sim 0.35-0.45$ in the r-magnitude\footnote{Assuming $M_r = -18.5$ and SDSS depth $m_r = 22.70$, \url{https://www.sdss.org/dr14/imaging/other_info/}}. However, more and more large scale surveys (such as SDSS) are now reaching these depths and include more dwarf galaxies, which means that the statistical basis for studying dwarf galaxies is becoming better. Furthermore, since dwarf galaxies constitute the first link in the chain of hierarchical structure formation theory, they constitute an invaluable source in figuring out the full galaxy formation and evolution puzzle.

Furthering their importance, most dwarf galaxies are believed to host intermediate-mass black holes \citep[e.g ][IMBHs; M$_{BH}\sim 10^{2}-10^6$ M$_\odot$]{Moran2014, Silk2017} -- a characteristic that has been studied in more detail in a number of papers; \citet{Barth2004} examined the host galaxy properties and the IMBH properties in the POX 52 galaxy. \citet{Reines2013} examined dwarf galaxies with optical signatures of active massive black holes. \citet{Sartori2015} searched for IMBHs using mid-IR and optical data while \citet{Baldassare2015} looked at the core region of RGG 118 and could infer an IMBH from the kinematics. 
Since IMBHs are the root of the super-massive black holes (SMBH) either through acting as a seed of gas accretion or merging of several IMBHs \citep[e.g ][]{Micic2007}, observing IMBHs during these phases (i.e AGN phase) can shed light on  conditions required for IMBH growth.

There are several mechanisms thought to trigger AGN activity. Merging or harassing galaxies are effective ways of accreting inter-stellar medium (ISM) or removing angular momentum from native gas reservoirs \citep{Miller2003, Sabater2013, Gordon2018, Ellison2019}, and the influx of material to the central regions of galaxies can then trigger AGN activity.  

Other AGN triggers include environmental effects \citep{Kauffmann2004}, where for example cooling gas from cluster cores accrete onto the central galaxies or the intergalactic medium compressing and shocking gas within a galaxy and driving the gas towards the core. Complicating this picture are observations that there might be a time delay between interactions and the onset of AGN activity\citep[e.g][]{Pimbblet2013}, which means that the current environment of a galaxy may not represent the environment that triggered the AGN activity. 

The effect of the environment on a single galaxy can be analysed from detailed and focused observations, but such an undertaking is not feasible for a large scale survey containing thousands of objects. However, several methods exist to quantify the environment of galaxies \citep[for a review, see][]{Muldrew2012}, which are more suitable for a study like this. For example, \citet{Miller2003} calculated a galactic density using the 10th nearest neighbour as the shell edge while \citet{Baldry2006} used the 4th and 5th nearest neighbour. \citet{Sabater2015} calculated a tidal estimator that traced the relation between tidal forces exerted by companions and the internal binding force of a galaxy. 

These methods all attempt to quantify the environment, but they all have different strengths and weaknesses. Some are better at describing the local galactic environment, others are better for the group/cluster environment while some are better for the immediate environment (i.e whether a close neighbour exerts strong influence or not). 

Even the task of identifying AGNs is not straightforward since they have different signatures in different wavelength regimes. In this work, spectroscopic data from the Sloan Digital Sky Survey SDSS will be used and two different AGN selection methods are utilised. Since this work is based on SDSS data, optical diagnostic diagrams are used. The first one is the common Baldwin, Phillips, \& Terlevich (BPT) diagram \citep{BPT} with the \citet{Kewley2001, Kauffmann2004} criteria for AGN. BPT takes advantage of the fact that different excitation mechanisms have different emission line fingerprints.

The second diagnostic is the less common WHAN diagram \citep{Cid2010, Cid2011}. WHAN utilises the equivalent width, $W_\lambda$, of H$\alpha$ and the [NII]/H$\alpha$ line ratio and thus covers the same wavelength regime as BPT. The WHAN diagram was developed as a response to the BPT since BPT leaves a large population of emission line galaxies (ELG) unclassified in SDSS data. The advantage is that it recovers most things that the BPT does, but it also gains the weaker AGNs. Both methods will be discussed further in section \ref{sec:classification}

Whether environment quenches AGN, triggers AGN, or has no effect, is unclear when it comes to dwarf galaxies. The broad goals of this work are to therefore determine the environment of dwarf galaxies with AGN characteristics and construct arguments based on these environmental measures on how such dwarf galaxies with AGN trigger and evolve. The environmental analysis consists of the 10th nearest neighbour (10NN) method and the velocity difference to nearest neighbours ($\Delta v_{NN}$), and the distributions for each sample is then compared to non-AGN galaxies using two-sample Kolgomorov-Smirnov tests.

This paper is structured as follows: Section~\ref{sec:dataMethods} contains details about the data and methods used. Section ~\ref{sec:analysis} includes the analysis and interpretation of the results and Section~\ref{sec:discussion} has discussions on the findings. Conclusions and a summary is found in Section~\ref{sec:conclusions}. This study assumes  a $\Lambda$-CDM Universe with $H_0=70$ and $\Omega_{m_0}=0.3$.

\section{Data and methods}
\label{sec:dataMethods}
This section describes the data used and the cuts made to classify dwarf galaxies and the diagnostics used to select AGN in that sample.

The selection criteria can be summarised as the following:
Low mass galaxies: $M_* \leq 5\times10^{9}M_\odot$,  $\sigma \leq 100 km/s$, and completeness corrections. BPT galaxies follow the classification in \cite{Kewley2001} while the WHAN AGN selection requires log([N II]/H$_\alpha) \geq -0.4$ and $W_{H_\alpha} \geq 3$Å.
The environment analysis involves two methods; distance to the 10th nearest neighbour and velocity difference to the nearest angular separated galaxy. 
\subsection{Data and sample selection}
\label{sec:data}
The data used for identifying dwarf galaxies and AGN is from the NASA-Sloan Atlas (NSA) catalogue. This catalogue is constructed by using several catalogues; Sources are found from a combination of SDSS DR8 \citep{York2000, Aihara2011}, NASA/IPAC Extragalactic Database\footnote{The NASA/IPAC Extragalactic Database (NED) is funded by the National Aeronautics and Space Administration and operated by the California Institute of Technology.},  Six-degree Field Galaxy Redshift Survey, Two-degree Field Galaxy Redshift Survey, ZCAT and ALFALFA catalogues. Spectroscopic measurements (e.g line fluxes) are performed on SDSS spectra while all catalogues are used to determine redshifts. The final NSA catalogue contains extragalactic sources to a high completeness to $z<0.05$, which there are 145,155 of. 

The detection and de-blending technique for the photometry analysis is described in \citet{Blanton2011}. It is in spirit based on the SDSS photometric pipeline \citep{Lupton2001}, but there are differences in the way objects are deblended and use r-band templates for all bands\footnote{For a more in-depth summary, see \url{http://nsatlas.org/documentation}}.
Furthermore, the sources in NSA are only included if they are matched to a spectroscopy survey. Not all sources have SDSS spectroscopy, but the ones that do, have had their spectra remeasured by \citet{Renbin2011} using an improved calibration, which affects small equivalent width lines  making this catalogue well suited for classification diagrams based on emission line ratios and equivalent widths.

Furthermore, since dwarf galaxies tend to have weaker emission, the better measurements (i.e higher signal-to-noise) of spectroscopic data make this catalogue preferable to others for this study. Another argument for this catalogue is the stricter significance in SDSS in r-band images on splitting 'child' objects from 'parent' objects -- basically when an algorithm decides that a source is two objects rather than one. For dwarf galaxies, it means fewer false positives making the dwarf galaxy sample more robust, although there is a risk of large galaxies 'absorbing' small and weak ones.

Low mass galaxies are selected by imposing a stellar mass limit of $\si{M_* \leq 5 \times 10^{9} M_\odot}$ and velocity dispersion $\sigma \leq \si{100\kilo\meter\per\s}$. This follows similar limitations as other work in the field (e.g \citealt{Reines2013} and \citealt{Penny2016, Penny2018}) and corresponds roughly to the stellar mass of the LMC. The masses in NSA is given in units of M$_\odot h^{-2}$, and while other studies assume $h\approx0.70$ \citep{Reines2013, Hainline2016, Baldassare2018}, we have assumed $h=1$ for galaxy masses despite the cosmology assumed. The analysis and results in Section~\ref{sec:analysis} and Table~\ref{tab:KStests} have been analysed using both values, and no significant difference is found. Therefore, the choice of $h=1$ remains unchanged. The effect on sample sizes can be seen in Table~\ref{tab:samples}.

\begin{figure*}
	\centering
	\includegraphics[width=\linewidth]{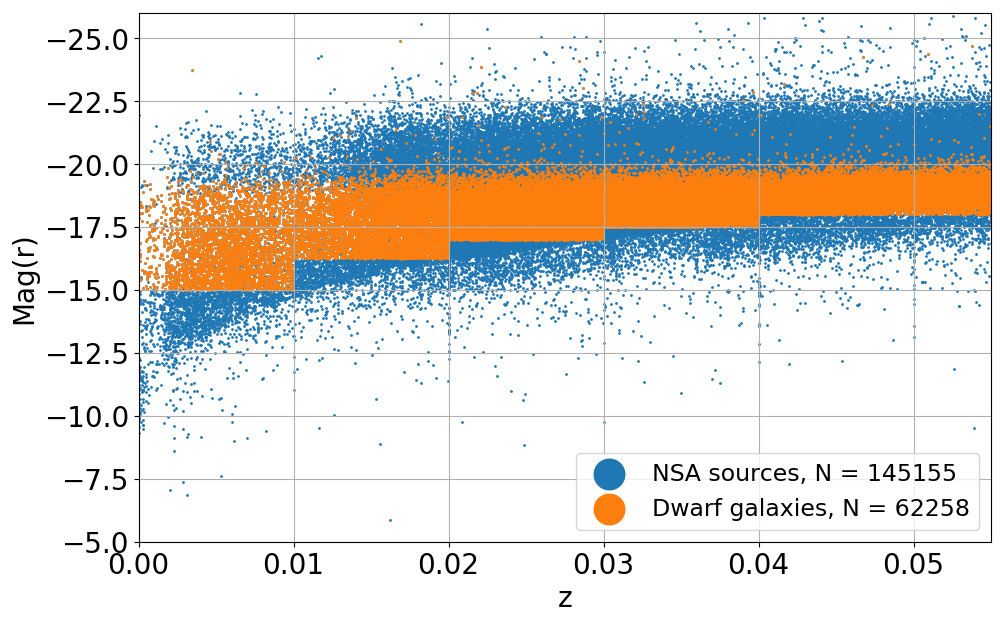}
	\caption{Magnitude versus redshift plot. The blue data points are all galaxies in NSA. The orange data points are low-mass galaxies (as defined in Section~\ref{sec:data}). There are clear magnitude edges in different redshift intervals, which is due to the completeness selection.}
	\label{fig:MvsZAll}
\end{figure*}

From inspection of $z$ vs $r$-magnitude (see Figure~\ref{fig:MvsZAll}), upper limits for several redshifts bins are imposed for the sake of completeness. The specific redshift bins and their corresponding magnitude-cuts can be seen in Table~\ref{tab:completeness}.  Using both the low mass galaxy criteria and the completeness restrictions, the sample size is reduced to 62,258 objects. This constitutes the parent sample from which further analysis is carried out.

\begin{table}
	\centering
	\caption{Completeness selection intervals.}
	\label{tab:completeness}
	\begin{tabular}{C{2cm}C{1cm}}	
	    \hline
		z & $\leq M_r$ \\
		\hline
		$0.00 \leq z < 0.01$ & -15.0 \\
		$0.01 \leq z < 0.02$ & -16.0 \\
		$0.02 \leq z < 0.03$ & -17.0 \\
		$0.03 \leq z < 0.04$ & -17.5 \\
		$z \geq 0.04$ & -18.0 		
	\end{tabular}
\end{table}

For the environmental analysis, the NSA catalogue is also used. The only interesting properties of the neighbour galaxies are their positions and redshift. The full number of sources is then 145,155 and all objects contain coordinates and redshifts from spectroscopy. The environmental analysis will be described in detail in Section~\ref{sec:envest}.

\subsection{Classification diagrams}
\label{sec:classification}
Two AGN selection methods are employed: The familiar BPT diagram \citep{BPT, Kewley2001, Kauffmann2003, Kauffmann2004} and the lesser-used WHAN \citep{Cid2010, Cid2011}. They are used both in conjunction and in parallel since they both have different strength and weaknesses. The diagnostics are used on the dwarf galaxy sample consisting of 62,258 objects. Below is a more detailed description of each classification scheme and an overview of the numbers can be found in Table~\ref{tab:samples}.

\begin{table}
	\centering
	\caption{Number of galaxies depending on choice of $h$. While the number of galaxies decreases with decreasing $h$, the results described in Section~\ref{sec:analysis} do not change.}
	\label{tab:samples}
	\begin{tabular}{lrrrrrr}	
	    \hline
		$h$ & Dwarfs & NOT & BPT & WHAN & AND & OR \\
		\hline
        $1.00$ &  62,258 & 55,643 & 387 & 4,323 & 228 & 4,476\\
    	$0.73$ & 43,774 & 41,341 & 124 & 1,399 & 62 & 1,461	\\
    	$0.70$ & 41,289 & 39,189 & 102 & 1,182 & 47 & 1,237	\\
	\end{tabular}
\end{table}

\subsubsection{BPT diagram}
\begin{figure}
	\includegraphics[width=\columnwidth]{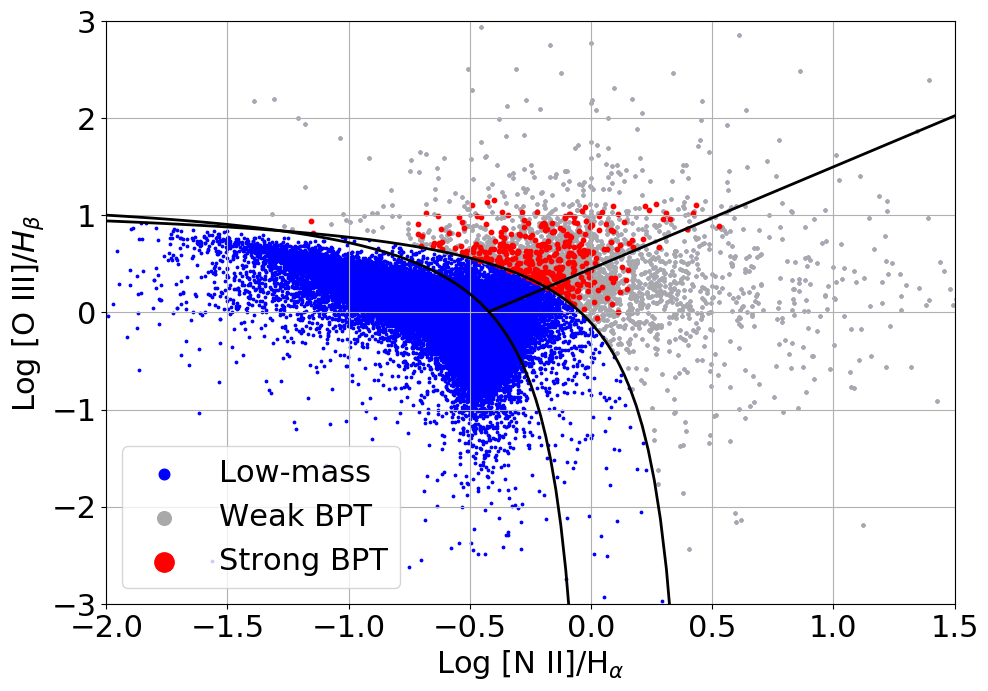}
	\caption{BPT diagram. The solid black lines are follow the \citet{Kewley2001, Kewley2006} classification diagram. However, no distinction is made between Seyfert and LINERS, and only pure AGNs are included in this sample, thus following the \citet{Kewley2001} classification. Three samples are plotted. The blue dots are all the low-mass galaxies in the NSA catalog. The red dots are the BPT-selected galaxies with S/N$_{\text{ratio}}>3/\sqrt{2}$ on  both emission lines ratios. The 'weak' BPT are galaxies with S/N$_{\text{ratio}}<3/\sqrt{2}$. Especially H$_\beta$ is responsible for classifying a BPT-selected galaxy as weak ($\approx 87.5\%$ of all dwarf BPT galaxies in this sample has S/N$_{H\beta}$ < 3).}
	\label{fig:BPTAll}
\end{figure}

The BPT diagram is used as one of the diagnostics to identify AGN. More specifically, the [N II]$\lambda 6584$/H$_\alpha$ vs. [O III]$\lambda 5007$/H$_\beta$ line ratios are used and follow the \citet{Kewley2001} distinction between composite star-forming galaxies and pure AGNs. While massive composite galaxies do include AGNs, too, we are uncertain of the intepretation in the low mass regime.
This division yields 2,644 objects. However, requiring a S/N $\geq 3$ on the 4 emission lines reduces this number to 296 -- a $\sim$88.8\% rejection rate.

The primary reason for rejection is due to the low S/N on H$_\beta$. 95.5\% of the 2,348 rejected BPT galaxies have a S/N$_{H_\beta} < 3$. As noted by \citet{Cid2010}, AGN galaxies have intrinsically low H$_\beta$ emission, which gives rise to low S/N measurements -- a problem that is exaggerated in dwarf galaxies because of their already weak signal. In \citet{Cid2010}, 53\% of their sample of emission line galaxies (ELGs) had weak measurements of H$_\beta$, which supports the notion that dwarf galaxies are particularly vulnerable to this effect.

Another approach is to require use the S/N$_{\text{ratio}}>3/\sqrt{2}$ instead. This means that if one emission line is well-determined but the other is not, it is not automatically rejected. This follows the same approach as e.g \citet{Juneau2014} and \citet{Trump2015}. Using this S/N cut yields 387 BPT galaxies and a $\sim85.4$\% rejection rate, and this is the sample used going forward.

BPT classification can also be performed using other line pairs such as [S II]$\lambda\lambda 6717,6731$/H$_\alpha$ and [OI]$\lambda 6300$/H$_\alpha$. However, they are also compared against [O III]$\lambda 5007$/H$_\beta$ and thus do not provide a way to bypass the low SNR on H$_\beta$. Therefore, these BPT diagrams are not chosen for further analysis in this work.

\subsubsection{WHAN diagram}
\begin{figure}
	\includegraphics[width=\columnwidth]{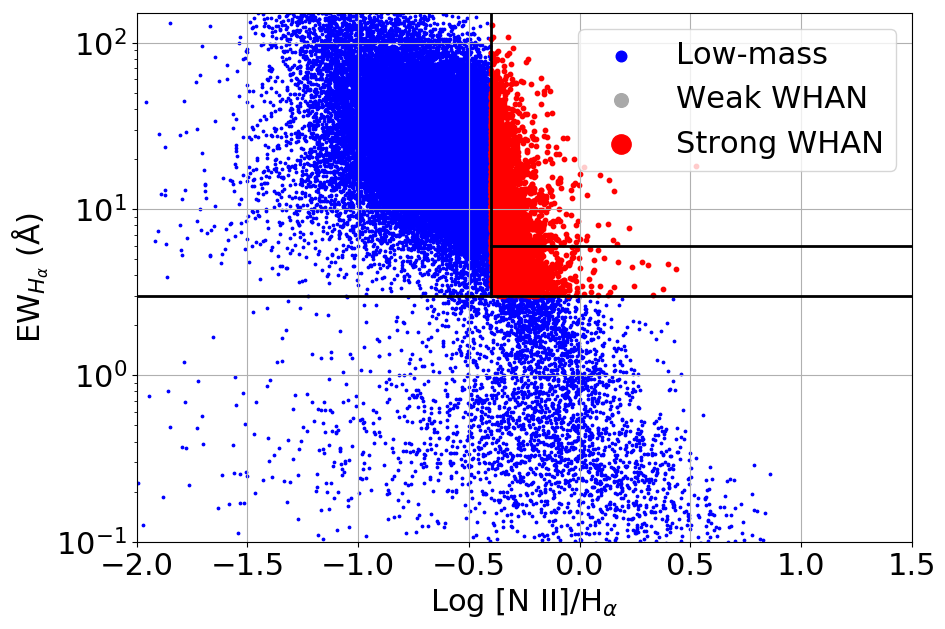}
	\caption{WHAN diagram \citep[For details, see][]{Cid2010, Cid2011}. The solid black lines mark the different regions (from top left, clock-wise); Star-forming, strong AGNs, weak AGNs, and retired galaxies. Both weak and strong AGNs are included in the sample and no distinction is made between them. The blue dots are all the low-mass galaxies in the NSA catalog. The red dots are the WHAN-selected galaxies with S/N$_{\text{ratio}}>3/\sqrt{2}$ on [N II]/H$_\alpha$. The 'weak' WHAN are galaxies with S/N$_{\text{ratio}}<3/\sqrt{2}$}
	\label{fig:WHANAll}
\end{figure}

The criteria for being classified as an AGN in the WHAN diagram follow \citet{Cid2011}; log([N II]/H$_\alpha) \geq -0.4$ and $W_{H_\alpha} \geq 3$Å. In the WHAN classification, there is a distinction between strong and weak AGN (\textit{weak} here meaning to be an indicator of energy output of the AGN and not low S/N like for the weak BPT classification). The used limit on $W_{H_\alpha}$ is such that both weak and strong AGN are included and no further distinction are made between them. This yields 4,323  objects. Using a S/N $\geq 3$ requirement of $H_\alpha$ and a S/N$_{\text{ratio}}>3/\sqrt{2}$ recovers 4,317 sources. This is the WHAN sample in onwards analysis.

\citet{Cid2010} suggest that the WHAN diagram is more suitable for selecting weaker AGNs -- especially ELGs -- compared to the BPT diagram. The BPT diagram is a very strict selection technique since it requires 4 emission lines of high quality. In fact, they argue that the choice of a strong (here meaning S/N$\geq 3$) H$_\beta$ biases against objects with low $W_\lambda$ and thus leaving out weaker AGN galaxies.
As the goal of this paper is to quantify the environment of dwarf galaxies hosting AGN, this makes the WHAN diagram an ideal selection method for our sample selection. 

\subsubsection{'AND', 'OR'  \& 'NOT' samples}
\begin{figure}
	\includegraphics[width=\columnwidth]{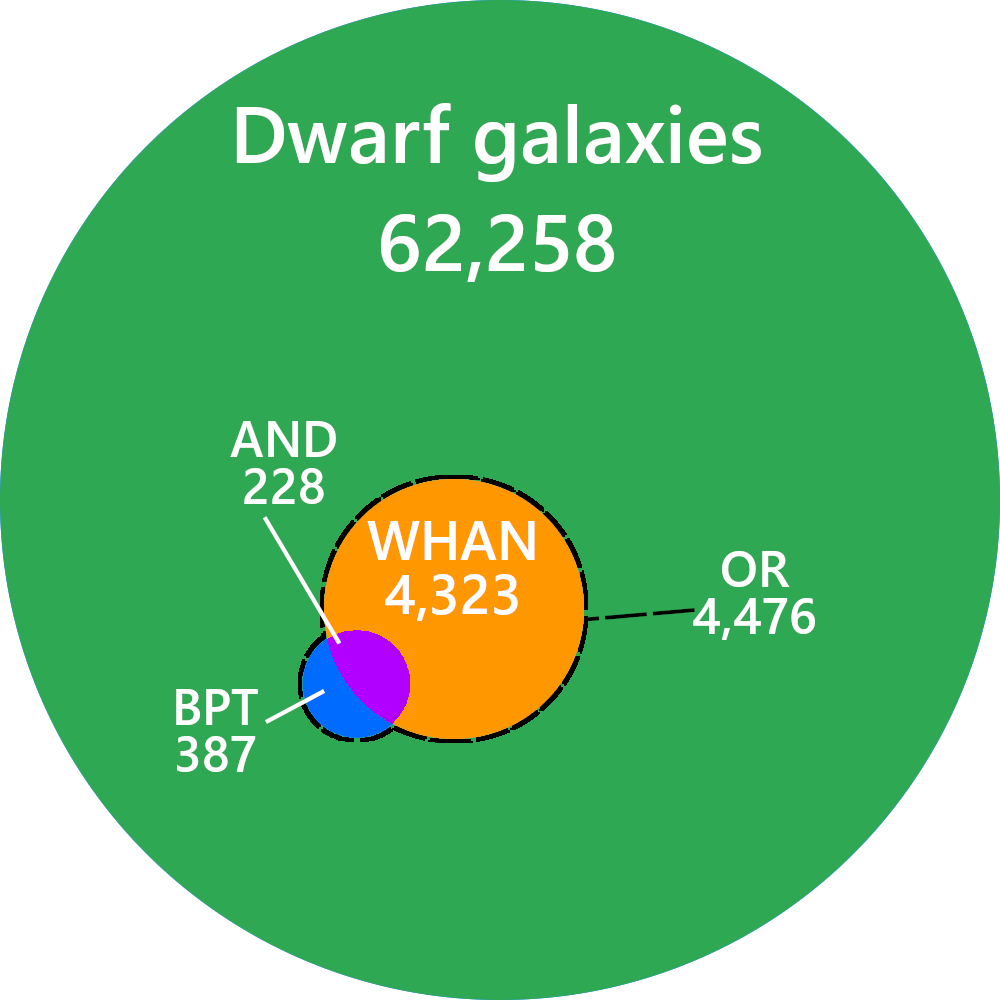}
	\caption{NSA Venn diagram showing the different selections.}
	\label{fig:NSA}
\end{figure}

As mentioned in Section~\ref{sec:introduction}, the two diagrams are used both separately and in conjunction with each other. Two further samples are made from the BPT and WHAN samples: \textit{'AND'} and \textit{'OR'}. The \textit{'AND'} sample is comprised of galaxies that fulfil \textit{both} the BPT and WHAN selection criteria while \textit{'OR'}-selected galaxies fulfil \textit{either}. The size of the samples are 228 and 4,476, respectively. 

Additionally, dwarf galaxies that does not appear in either non-S/N-corrected sample are labelled \textit{'NOT'}. This is the largest subsample and comprises 55,643 objects. In \textit{'NOT'}, galaxies with either a BPT or WHAN AGN classification (before correcting for low S/N and are thus not considered AGNs in this study). They are not included because the sample size is sufficiently large without them -- even if they were included, it would only change the sample size by less than $\sim$4\% and excluding them makes the \textit{'NOT'} sample more robust because only onjects with clear classification is included.

\subsection{Environment estimation}
\label{sec:envest}
There are two different properties of the environment that this study attempts to examine: The density of the local environment and the recent interaction history of AGN galaxies where the local environment is to be understood as the area of the group or cluster that the dwarf galaxy is situated in. Though both properties are not straight-forward to quantify, there are a number of methods to infer them \citep[for a discussion of these, see][]{Muldrew2012}. 

One method to infer the density of the environment is the projected distance to the 10th nearest neighbour (10NN) while the recent interaction history can be inferred from the velocity difference to the nearest angular separated neighbour ($\Delta v_{NN}$). Throughout this study, the environment inferred from 10NN is often referred to as the local environment, and from $\Delta v_{NN}$, the environment is referred to as the immediate environment. 
While other studies describe the local environment by the galaxy surface density, translating $r_{10}$ to galaxy surface density is straightforward through the equation $\Sigma_{10} = \frac{N}{\pi r_{10}}$. Therefore, the use of $r_{10}$ is as good as $\Sigma_{10}$.

Both methods only consider galaxies within $\pm$1,000 km/s, which is slightly higher than the average galaxy cluster velocity dispersions \citep[see e.g ][]{Bilton2018}. This is to ensure that galaxies are only neighbours if they are close spatially (i.e member of the same group or cluster) and not just angularly close.  \citet{Muldrew2012} remarks that a nearest neighbour approach is a better measure of the local density compared to cluster density, and a higher \textit{n}th separated neighbour smooths out local variances. Smoothing out local variances is desirable for estimating the local environment in general, but local variances are exactly what is important for immediate environments. 

\section{Analysis}
\label{sec:analysis}
This section contains the statistical analysis of differences between the subsamples. The neighbourhood parameters (10NN and $\Delta v_{NN}$) will be looked at with a Monte Carlo Kolmogorov-Smirnov (KS) test procedure while other properties such as stellar mass and redshift will receive a short statistical rundown. A visual inspection is also carried out on the galaxies fulfilling both the WHAN and BPT criteria (i.e the \textit{'AND'} subsample) and compared to a similar sized subsample from the \textit{'NOT'} subsample. 

\subsection{KS-testing}
\label{sec:kstesting}
\begin{table}
	\centering
	\caption{p-values of respectively 10NN and $\Delta v_{NN}$ 2-sided KS tests. Each row has the subsample in the leftmost column as the subsample to be compared against a control sample from a subsample given by the column name. E.g, the test in row 1, column 2 is found from 152 random galaxies from all low mass galaxies and a matching galaxy (in mass, colour, and redshift) sample is found for each element from the BPT subsample. 'wBPT' is short for 'weak BPT'.}
	\label{tab:KStests}
\begin{tabular}{l r r r r r r r}
\hline
	10NN & \multicolumn{1}{l}{\textbf{All}} & \multicolumn{1}{l}{\textbf{BPT}} & \multicolumn{1}{l}{\textbf{WHAN}} & \multicolumn{1}{l}{\textbf{AND}} & \multicolumn{1}{l}{\textbf{OR}} & \multicolumn{1}{l}{\textbf{NOT}} & \multicolumn{1}{l}{\textbf{wBPT}} \\
		\hline 
	\textbf{All}  & 0.52    & 0.26    & 0.47  & 0.14    & 0.46   & 0.54    & 0.00  \\
	\textbf{BPT}  & 0.20    & 0.57    & 0.40  & 0.43    & 0.40   & 0.21    & 0.00  \\
	\textbf{WHAN} & 0.52    & 0.27    & 0.55  & 0.12    & 0.53   & 0.50    & 0.00  \\
	\textbf{AND}  & 0.14    & 0.51    & 0.25  & 0.53    & 0.26   & 0.13    & 0.00  \\
	\textbf{OR}   & 0.50    & 0.27    & 0.54  & 0.13    & 0.56   & 0.49    & 0.00  \\
	\textbf{NOT}  & 0.55    & 0.28    & 0.44  & 0.14    & 0.44   & 0.54    & 0.00  \\
	\textbf{wBPT} & 0.17    & 0.00    & 0.01  & 0.00    & 0.01   & 0.20    & 0.56 \\ \hline
\end{tabular}
\linebreak \linebreak \linebreak
\begin{tabular}{lrrrrrrr}
	\hline
	$\Delta v_{\text{NN}}$ & \multicolumn{1}{l}{\textbf{All}} & \multicolumn{1}{l}{\textbf{BPT}} & \multicolumn{1}{l}{\textbf{WHAN}} & \multicolumn{1}{l}{\textbf{AND}} & \multicolumn{1}{l}{\textbf{OR}} & \multicolumn{1}{l}{\textbf{NOT}} & \multicolumn{1}{l}{\textbf{wBPT}} \\
	\hline
	\textbf{All}  & 0.53  & 0.31    & 0.46  & 0.40    & 0.46   & 0.53    & 0.39  \\
	\textbf{BPT}  & 0.44  & 0.56    & 0.41  & 0.46    & 0.43   & 0.44    & 0.29  \\
	\textbf{WHAN} & 0.49  & 0.33    & 0.52  & 0.25    & 0.53   & 0.48    & 0.38  \\
	\textbf{AND}  & 0.32  & 0.49    & 0.28  & 0.55    & 0.29   & 0.31    & 0.21  \\
	\textbf{OR}   & 0.51  & 0.34    & 0.54  & 0.26    & 0.52   & 0.50    & 0.36  \\
	\textbf{NOT}  & 0.51  & 0.28    & 0.43  & 0.39    & 0.45   & 0.53    & 0.37  \\
	\textbf{wBPT} & 0.45 & 0.34    & 0.37  & 0.22    & 0.38   & 0.47    & 0.53    \\
	\hline
\end{tabular}
\end{table}

\begin{figure}
	\includegraphics[width=\columnwidth]{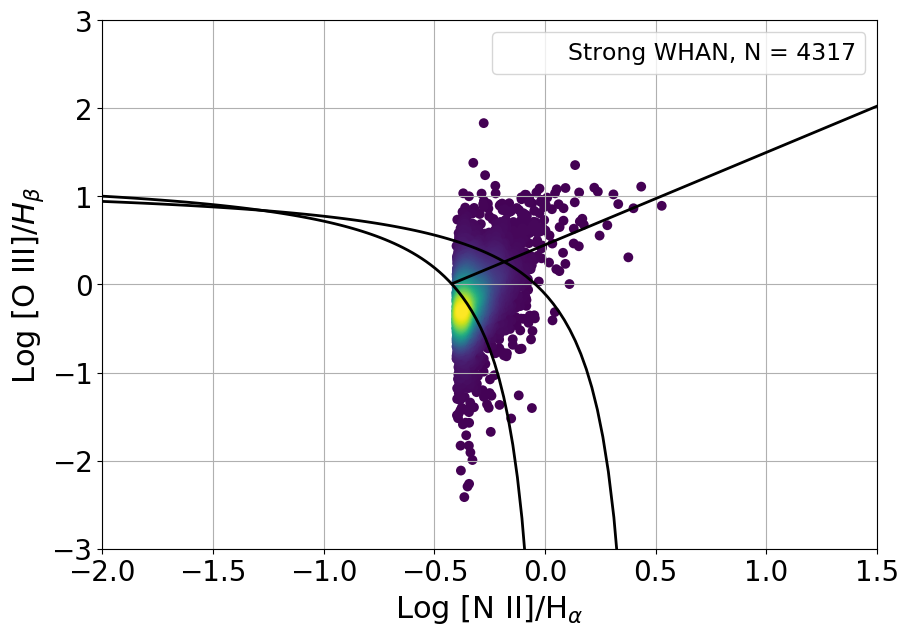}
	\caption{BPT diagram with WHAN selected galaxies. The dots are colour-coded by their relative point density. The majority of the WHAN selected galaxies would have been classified as star-forming or composite SF/AGN using the BPT classification scheme.}
	\label{fig:WHANinBPT}
	
	\includegraphics[width=\columnwidth]{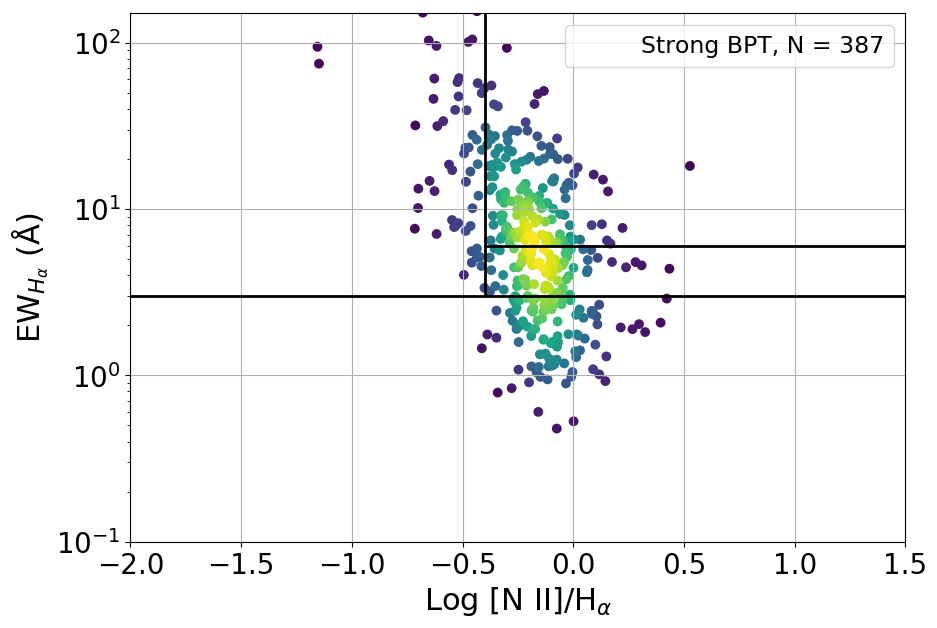}
	\caption{WHAN diagram with BPT selected galaxies. The dots are colour-coded by their relative point density. The majority of BPT selected galaxies (\textit{'AND'}-selected - N = 195) are considered strong AGNs in the WHAN diagram while the non-AGN WHAN-classified galaxies are roughly evenly split between retired galaxies and star forming ones.}
	\label{fig:BPTinWHAN}
\end{figure}

\begin{figure}
	\includegraphics[width=\columnwidth]{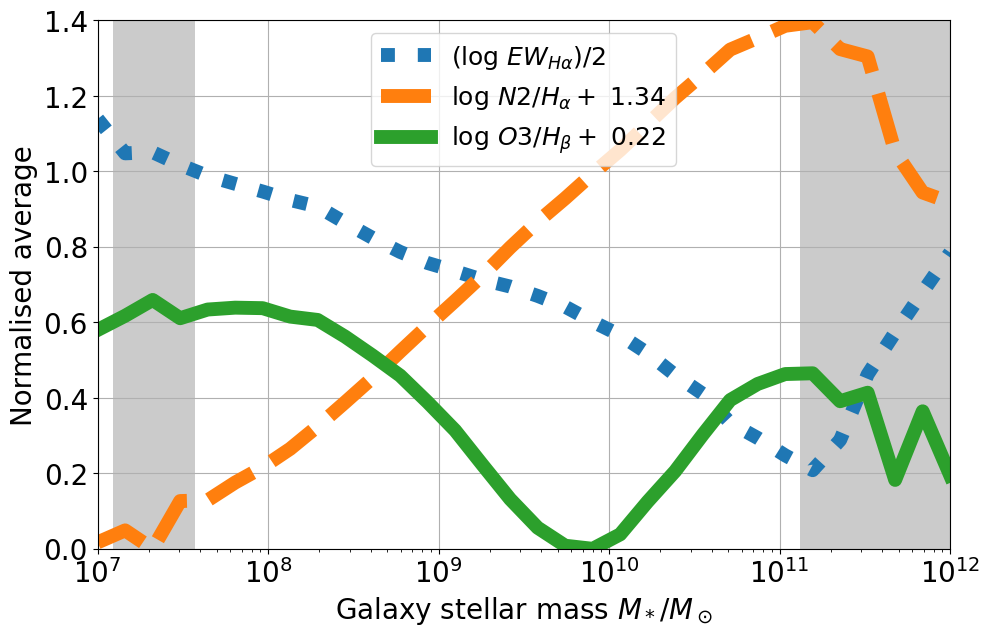}
	\caption{Average values of emission line ratios and EW$_{H_\alpha}$ function of mass. The log ratio values are shifted to be in the same area while the EW is log and then scaled by 0.5. The data consists of 32 linear log scale mass bins and the bins with less than 300 galaxies are shaded in grey. }
	\label{fig:allAvgs}
	
\end{figure}

\begin{figure}
	\texttt{}\includegraphics[width=\columnwidth]{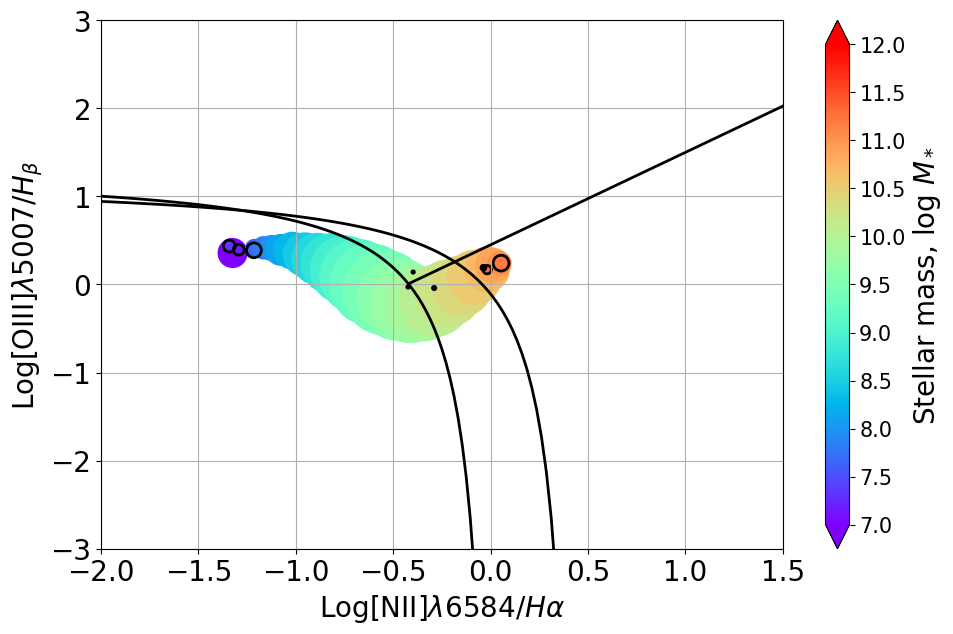}
	\label{fig:BPTMassAvgs}
	
	\includegraphics[width=\columnwidth]{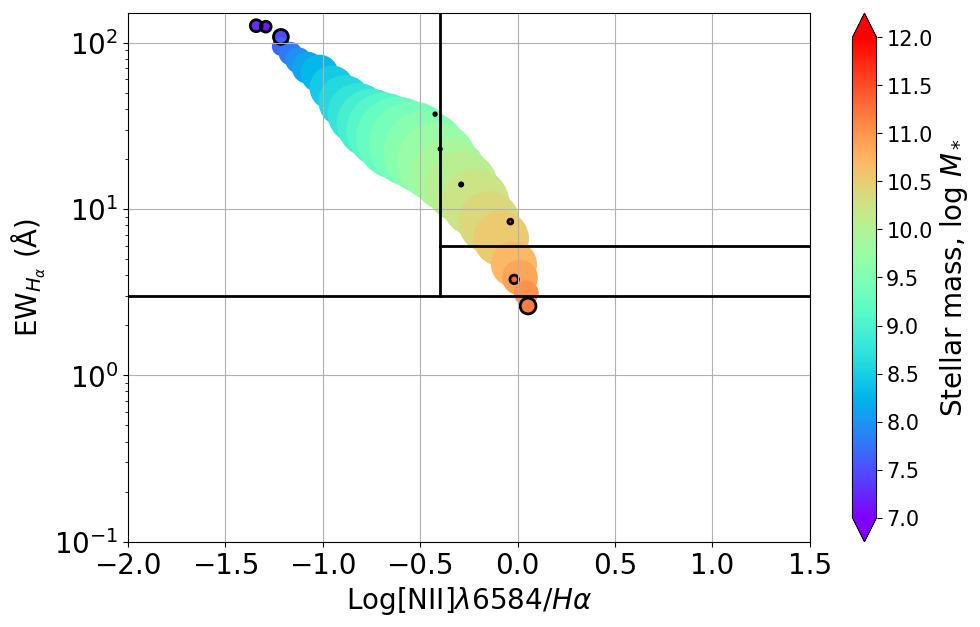}
	\caption{BPT and WHAN diagram showing mass trends. Each data point is the average values in 32 different mass bins and the size of the dot is scaled by the number of galaxies in that bin. Each bin has at least 300 galaxies in them unless surrounded by a black edge and otherwise contains between 312 and 11,581 galaxies.}
	\label{fig:WHANMassAvgs}
\end{figure}
To quantify the difference of environment between different subsamples, two-sample Kolmogorov Smirnov (KS) tests are carried out. It is a test of whether or not two samples come from the same parent distribution -- for example whether the distribution of the distance to the 10th nearest neighbour of the 'BPT' sample is the same as the distribution of the 'NOT' sample. Though two samples of different sizes can be used, the input sample sizes are scaled to 152 elements. 152 is the number of objects in the smallest subsample (WHAN AGNs that are rejected in BPT because of low S/N).

Each KS test is iterated 1,000 times, each time with 152 different random elements from the subsamples listed in Table~\ref{tab:KStests}. Next, a comparison sample is found from another subsample (although each subsample is also tested against itself) where a matching galaxy is found for each of the 152 in the original subsample. The matching critera involves mass, redshift, and colour. The critera are:
\begin{enumerate}
    \item $\|1-M_{*}/ M_{*,\text{AGN}} \| \leq 0.20$
    \item $\|z_{\text{AGN}} - z \| \leq 0.01$
    \item $\|(u_\text{AGN}-r_\text{AGN}) - (u-r)\| \leq 0.4$
\end{enumerate}
The matching criteria are similar to that of \citet{Cheung2015} but with stricter limits on mass and redshift. The masses of the galaxies in this sample are more similar than the galaxies in of \citet{Cheung2015} and are more numerous which allows for more strict criteria without eliminating all possible matches. 

A stricter redshift interval is necessary due to that fact that the SDSS fiber covers different fraction of galaxies at different redshift -- an effect that is very pronounced at lower redshifts. E.g, at $z=0.005$, the 3" fiber covers 0.3 kpc while at $z=0.055$, it covers 3.3 kpc. Thus, it is not the same reagion of each galaxy that is examined with redshift. A discussion of this effect can be found in Section~\ref{sec:fiberbias}.

This follows the same methodology as in other papers such as \citet{Penny2016}. That study's sample size is smaller (39) since it is drawn from the smaller MaNGA survey. Therefore, the statistics are not directly comparable because the p-value from KS-testing changes with sample size (decreases with larger sample size). They are sufficiently similar to allow for adaptation of the method. The values shown in Table~\ref{tab:KStests} shows the average p-values of these iterations.

\subsection{BPT and WHAN comparison}
To compare the two selection methods, the AGN-subsamples are classified in the other's diagnostic diagram (see Figure~\ref{fig:WHANinBPT} and \ref{fig:BPTinWHAN}). While BPT galaxies tend to be in the AGN part of the WHAN diagram, WHAN galaxies are mostly in the star-forming or composite region in the BPT diagram. Interestingly, $\sim$53\% of WHAN selected galaxies have S/N$_{H_\beta, [O III]} \geq 3$, which is the same fraction as \citet{Cid2010} found for all galaxies. This means that they are more robustly classified in the BPT diagram than the initial (i.e before SNR rejection) BPT galaxies. The BPT diagram classifies the majority of WHAN galaxies as only star-forming but with a significant number of composite galaxies.

An interesting finding is that the emission line ratios and equivalent width of $H_\alpha$ all have clear mass trends. Towards lower stellar mass galaxies, [N II]/$H_\alpha$ decreases while EW$_{H_\alpha}$ increases. This means that galaxies move towards the upper left corner of the WHAN diagram -- deep in the star formation region. This is in agreement with the literature on dwarf galaxies that they are very star forming \citep{Kauffmann2004, Yang2007, Geha2012}. Towards higher masses, the average EW$_{H_\alpha}$ drops to below 3 Å, which helps explain why the WHAN AGN fraction peaks around $M_* = 10^{10}$ (see Figure~\ref{fig:AGNFraction} for a visualisation). 

In the BPT diagram towards lower masses, the trend of [N II]/$H_\alpha$ moves the galaxies away from the vertical cut-off for AGN/LINER classification, and the [O III]/$H_\beta$ trend for $M_* \leq 10^9$ is declining while it increases afterwards. In BPT, whenever [N II]/$H_\alpha \gtrsim - 0.1$, galaxies are classified as either composite or pure AGNs. This condition is met for the average [N II]/$H_\alpha$ for galaxies $2\times10^{10} M_\odot\geq M_* \geq 2\times10^{11} M_\odot$ possibly explaining the BPT AGN fraction peak around $M_* \sim 10^{11}$.

\subsection{Local neighbourhoods of dwarf AGNs, 10NN}
From the KS-testing, it appears that there are no discernible differences between the distances to the 10th nearest neighbours of any of the subsamples. This means that the density of the environment does not seem to affect AGN activity in dwarf galaxies, and the implications will be discussed further in \ref{sec:envdiscussion}. Figure~\ref{fig:dist10thAll} shows the 10NN distribution BPT, WHAN and NOT and similar figures for the \textit{AND}, \textit{OR}, and weak BPT and \textit{NOT} samples can be found in the appendix. The statistics can be found in Table~\ref{tab:KStests}.

The average projected separations are between $d_p = 3.7 - 4.3$ Mpc with $\sigma = 2.0 - 2.2$ Mpc, which further shows that the distributions are indiscernible. The BPT and WHAN distributions tend to lie at the lower end of both intervals (respectively, $3.7 \pm 2.0$ Mpc and $4.1 \pm 2.2$ Mpc) suggesting they do prefer denser environments compared to NOT galaxies ($4.3 \pm 2.2$ Mpc) though the KS statistics make this inconclusive.

\subsubsection{Weak galaxies}
\label{sec:weakKS}
The only subsample that shows a significant difference in distribution is BPT selected galaxies with low S/N and thus rejected as AGNs.
Though it is uncertain whether this subsample has AGN characteristics due to low S/N, this subsample will be referred to as 'weak BPT'. These galaxies will be discussed further in Section~\ref{sec:weakDiscussion}.

\subsection{Immediate neighbourhood of dwarf AGNs, $\Delta v_{NN}$}
\label{sec:dvAnalysis}
\begin{figure*}
	\includegraphics[width=\linewidth]{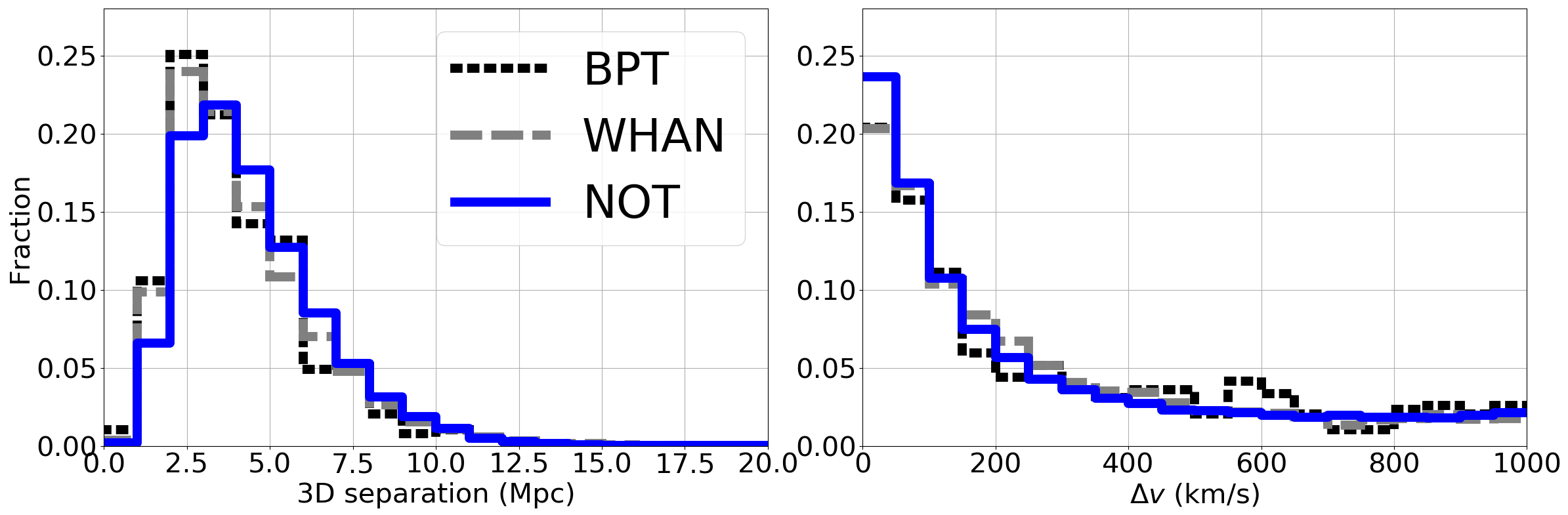}
	\caption{\textit{Left}: Projected spatial separation from the dwarf AGN galaxies to their 10th nearest neighbour and \textit{right}: The absolute velocity difference between dwarf AGN galaxies and their nearest 2D separated neighbour (within $\pm$ 1,000 km/s). Three samples are plotted: Black dotted are BPT-selected galaxies, grey dashed are WHAN-selected galaxies while blue solid are galaxies that appear in neither of the other samples. Generally, there are no discernable differences between the three distributions in either case. The BPT bump near 600 km/s is not statistically significant. See Section~\ref{sec:kstesting} and Table~\ref{tab:KStests} for statistics.}
	\label{fig:dist10thAll}
	
	\label{fig:DVHistAll}
\end{figure*}

Similarly to 10NN, this measure shows no discernible between the any of the subsamples -- even weak emission line galaxies. This seems to suggest that the velocity difference to a dwarf AGN galaxy's nearest neighbour is not deciding factor in its AGN activity. The distributions can be seen in Figure~\ref{fig:DVHistAll}. A notable anomaly/feature is an excess at around 600 km/s in the BPT distribution, but this 'bump' does not significantly affect the KS statistics. However, the bump does seem to make the BPT distribution have the highest average $\Delta$v. 

Overall, most galaxies tend to have a very small velocity differences to its nearest neighbour. There is no adjustment for the fact that the velocities are only in the line of sight, which partially explains the shape of the distribution.

\subsection{Visual inspection}
\label{sec:visinsp}
A visual inspection is carried out to look for any morphological disruptions. Such tidal interactions are not necessarily quantified by the two primary environment estimation methods and thus serves as a complementary qualitative method. Two subsamples are used: The \textit{'AND'} subsample and a simlar sized control sample from the \textit{'NOT'} subsample. The control subsample is comprised of galaxies that are matched in stellar mass, colour, and redshift to the AGN galaxies. The matching criteria are the same as in Section~\ref{sec:kstesting}. The purpose of this is to look for any obvious asymmetries or tidal interactions with neighbours.

\begin{figure*}
	\centering
	\includegraphics[width=\linewidth]{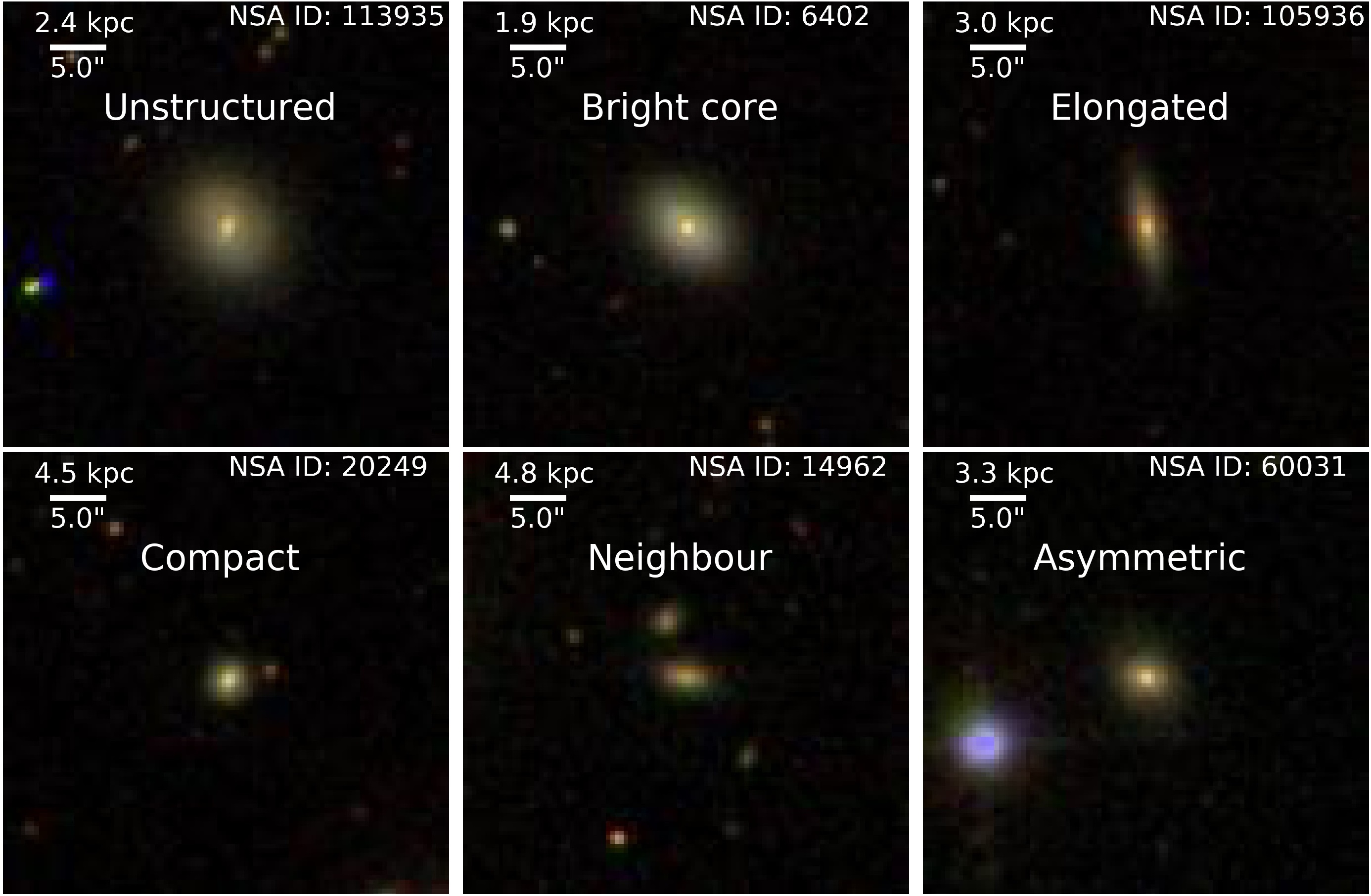}
	\caption{Example of cutouts of SDSS data. 6 different observational properties are shown here (only excluding spirals). Detailed information regarding visual inspection can be found in Section~\ref{sec:visinsp}}
	\label{fig:cutouts}
\end{figure*}

The images are 40'' by 40'' and from SDSS. They are characterised by a number of properties which will be explained below. Figure~\ref{fig:cutouts} showcases 4 of these properties in the different subsamples. A number of galaxies are rejected due to either appearing as a massive galaxy or observational artefacts. 
\begin{enumerate}
\item Unstructured. Does the galaxy lack any morphology or have any discernible structure, e.g spiral arms or dust lanes? 
\item Bright core. Does the galaxy have a concentrated peak in brightness at the centre?
\item Elongated. Is the galaxy flatter than roughly an $\sim$E6 galaxy?
\item Compact. The appearance of the galaxy is that only of a core and confined within $4''$
\item Spiral. Does the galaxy show clear spiral arms from either an angle or face-on? 
\item Neighbour. Does the galaxy have a neighbour in the image? A neighbour is defined as a source of roughly the same colour and brightness.
\item Asymmetric. Does the galaxy have asymmetric features such as a tidal tail, a warped appearance or unevenly distributed light.
\end{enumerate}

Item (i)-(v) are descriptive of the intrinsic properties of the galaxies whereas (vi) and (vii) can be used to infer properties about their environments. The numbers between the two samples are similar (within $\sim 6$ \%-points) in most aspects except frequency of bright cores and being compact.

The higher frequency of bright cores and compactness of AGNs can be explained by the intrinsic properties of AGNs: They are defined as having a high degree of radiation originating from the nucleus, which explains the bright cores. Furthermore, a galaxy with weak galactic emission or being at a high redshift with a relatively strong AGN results in only the core region  being visible thus appearing compact on the sky.

The neighbour numbers are somewhat comparable to \citet{Ellison2019} that found roughly 78 \% of non-AGN galaxies and 64 \% of AGN galaxies to be isolated and whereas the numbers for this study is  78 \% and 75 \% respectively. Tidal features (equivalent to asymmetries in this study) for AGNs in \citet{Ellison2019} are higher than the fraction in this study, but their numbers for non-AGNs are comparable to the control sample here. The sample size in this study is lower by a factor of $\sim 6$, though, and the approach here is not as meticulous as \citet{Ellison2019}, which may partially account for the difference.

Furthermore, it should be noted that mass and redshift distributions are different in \citet{Ellison2019}, so some differences are expected. More specifically, their sample goes to $z \simeq 0.25$ and only 3 out of 1,124 optical galaxies are low mass galaxies and 7 out of 254 mid-IR galaxies. Also, more massive galaxies are larger and will have a larger angular size on average than the dwarf galaxies, which would make tidal features easier to spot\footnote{A galaxy the size of the Milky Way (MW, $d \simeq 30$ kpc) at $z = 0.25$ will have the same angular size as an LMC-like galaxy ($d \simeq 4$ kpc) at $z=0.03$. Thus, most galaxies in \citet{Ellison2019} are more resolved than even the largest galaxies in this sample at $z=0.03$ -- yet the sample of this study even goes to $z=0.055$. For a MW-like to have the same angular size as an LMC-like galaxy, its redshift would have to be $z=0.48$}.

\begin{table}
	\centering
	\caption{Number (fraction) of galaxies showing visual properties in AGNs (\textit{'AND'} subsample) and a control sample.}
	\label{tab:visualInsp}
	\begin{tabular}{l  rlrl }	
	    \hline
		Parameter & \textbf{AGN (181)} & & \textbf{Control (192)} &\\
		\hline
		Structureless   & 156 & (86\%)    & 177 & (92\%)    \\
		Bright core     & 163 & (90\%)    & 81 & (42\%)     \\
		Elongated       & 61 & (34\%)     & 62 & (32\%)     \\
		Compact         & 76 & (42\%)     & 59 & (31\%)     \\   
		Spiral          & 17 &  (9\%)     & 18 &  (9\%)     \\
		Neighbour (N)   & 45 & (25\%)     & 43 & (22\%)     \\
		Asymmetric (AS) & 21 & (12\%)     & 19 & (10\%)     \\
		N+AS            & 7 & (4\%)       & 5 & (3\%)       \\
		\hline
	\end{tabular}   
\end{table}

\subsection{Other parameters}
\begin{figure}
	\includegraphics[width=\columnwidth]{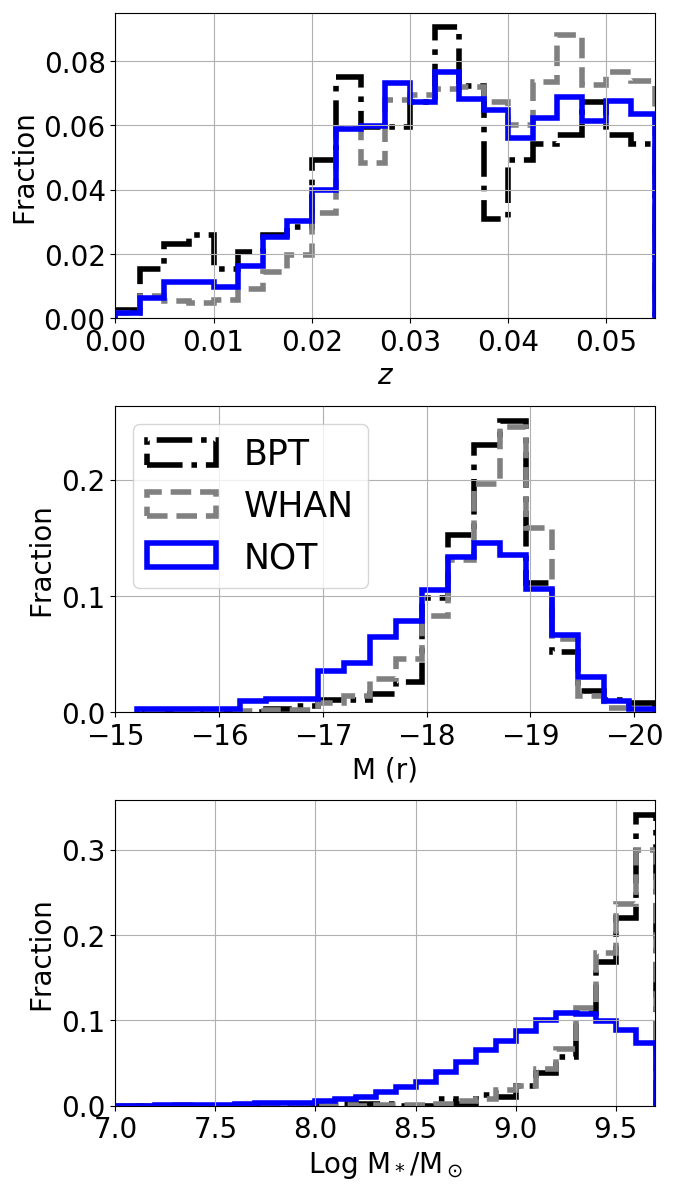}

	\caption{Mass, redshift, and magnitude distributions. The black dash-dotted distribution is BPT-selected galaxies, the grey dashed is WHAN-selected ones while the blue solid is the NOT selection. Regarding mass, AGN galaxies are increasingly common towards higher masses while the NOT galaxies peak around log 9.3 $M_\odot$. For redshift, WHAN and NOT galaxies follow almost the exact same trend, though WHAN has a slight excess at higher redshifts. BPT galaxies are slightly favoured at lower redshifts, but overall follows the same trend. Lastly, on magnitude, AGN galaxies are in general brighter than the NOT galaxies.}
	\label{fig:MHistAll}
	\label{fig:zHistAll}
	\label{fig:massHistAll}
	\end{figure}
\begin{figure}
	\includegraphics[width=\columnwidth]{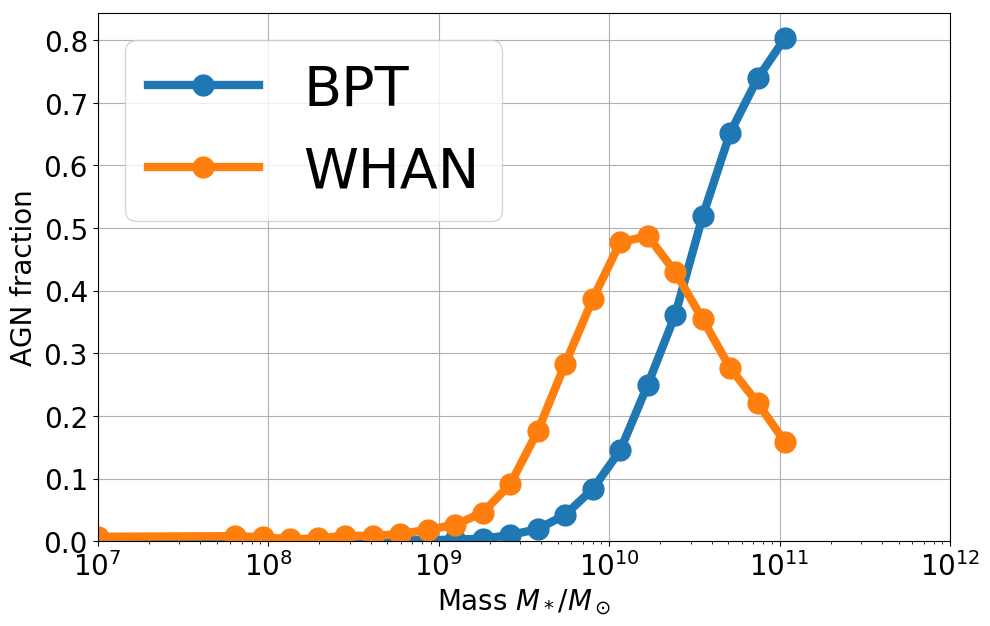}
	\caption{AGN fraction as function of mass. The fraction is calculated as the number of galaxies fulfilling the respective AGN criteria divided by the total number of galaxies in that mass bin that also fulfill the S/N criteria outlined in Section~\ref{sec:classification}. For high masses ($\geq 10^{11} M_\odot$), care has to be taken because of incomplete data, which is why there are no mass bins after $\sim 10^{11} M_*/M_\odot$ since a requirement is that there has to be more than 300 galaxies in one bin.}
	\label{fig:AGNFraction}
\end{figure}
The other parameters that are compared are: Stellar mass, $M_*$, redshift, $z$, and \textit{r}-magnitude. The conclusions from them are mostly used to check whether or not the samples behave as expected -- e.g higher AGN fraction at higher masses. 
The distributions are in Figure~\ref{fig:MHistAll}. 

The mass distribution is not surprising. AGNs are primarily found in the higher mass systems which agrees with e.g \citet{Miller2003, Kauffmann2004,Sabater2013}, and these distributions follow the same trend. BPT and WHAN are indiscernible with a very rapid rise from log$(M_*) \approx 9.2 M_\odot$ while \textit{'NOT'} plateaus around that mass and falls off steadily -- a trend that is partially caused by AGN distributions 'stealing' galaxies at these masses. 
The AGN fraction as a function of mass bin (Figure~\ref{fig:AGNFraction}) does reveal that the mass cut of $M_* \leq 5 \times 10^{9} M_\odot$ is where the AGN fraction starts changing the most in WHAN

The majority of the low mass galaxies are found at redshifts above $z=0.02$. Generally, all distributions follow the same pattern, although BPT galaxies seem to be found and lower redshifts whereas WHAN has more objects at higher redshifts. 

The magnitudes of the subsamples show that AGN galaxies tend to be brighter than non-AGN galaxies. There are quite a few parameters to untangle, though. Since the mass distribution of AGNs in this sample are shifted towards higher masses, it seems natural that the magnitudes are shifted accordingly. However, the masses are derived from the r-magnitudes \citep{Blanton2011}, which may mean that some of the luminosity from the AGN contributes to the stellar mass estimate. 
What it shows at least is that the subsamples behave as expected with active galaxies being more luminous than regular dwarf ones.

\section{Discussion}
\label{sec:discussion}
Overall, the apparent non-dependence on environment of dwarf galaxy AGN hosts found in this study is in line with existing literature \citep[e.g][although several of these studies find other properties that trend with environment like AGN colour or OIII strength]{Miller2003, Kauffmann2004, Padilla2010, Man2019}. It suggests that AGNs in dwarf galaxies react similarly to environment as regular galaxies. 

This non-uniqueness of dwarf galaxies is surprising since the gravitational potential, cold gas content \citep[e.g][]{Bradford2018}, and morphology  of dwarf galaxies are different to regular galaxies. \citet{Sabater2013} suggest the most important factor in fueling AGN activity is having a supply of gas to feed the core, and the cold gas content is more vulnerable in dwarf galaxies due to their shallow gravitational potentials. Dense local environments have a detrimental effect on the cold gas reservoirs by stripping and heating it while strong galaxy interactions can enhance AGN activity by perturbing the otherwise stable structures, though neither effects can be inferred from the results of this study..

The implications are that; a) environment plays an insignificant role on AGN activity regardless of host mass, or b) the environmental effect on AGN activity is either delayed or obfuscated such that the environment measurement methods do not probe the desired properties, or c) the selection methods for AGN cannot be applied directly to the low mass regime due to biases, or d) a mix of the above. 

\subsection{SDSS fiber aperture bias}
\label{sec:fiberbias}
\begin{figure}
	\centering
	\includegraphics[width=\linewidth]{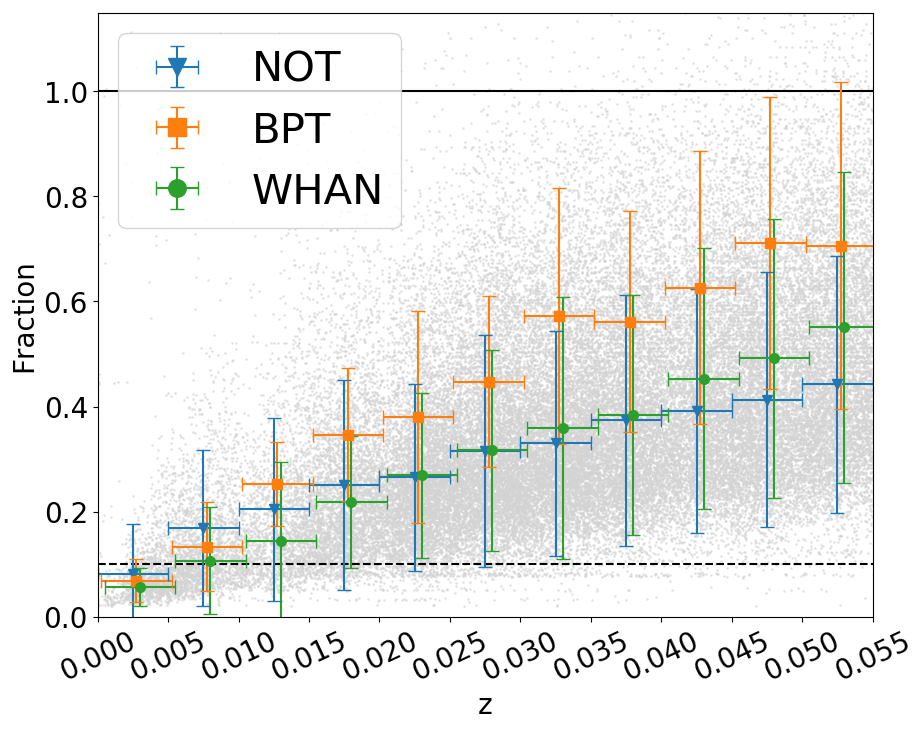}
	\caption{Plot of fraction of a galaxy covered by the SDSS fiber aperture as function of redshift. The size of a galaxy is taken to be its petrosian 90\% light radius, $R_{P90}$, and is core is defined as  $0.1 R_{P90}$. The grey dots are all dwarf galaxies overplotted with median of different subsamples. Errorbars show the interquartile range. The subsamples are split into redshift bins with $\Delta z=0.005$. The solid line at 1.0 equals the $R_{P90}$ while the dashed one at 0.1 is $0.1 R_{P90}$.}
	\label{fig:galcover}
\end{figure}

The SDSS fiber aperture does not cover the same fraction of a galaxy as a function of redshift, which can be seen from Figure~\ref{fig:galcover}. In the lowest redshift bin, the fiber does not even cover the whole core. This means that measured emission lines are affected by the redshift of the galaxy. Emission line flux will be left out if the aperture size covers less than the core while AGN signatures may be drowned out by SF emission for larger fractions \citep[e.g][]{Trump2015}. 

Although the galaxy cores are not resolved, if a core region is smaller than the fiber diameter, the total AGN flux will not be recovered which will lead to inaccurate emission line measurements thus making equivalent width methods such as WHAN inaccurate. The emission line ratios may not be affected though making the BPT diagnostic robust.

A test is performed to see if excluding galaxies whose core regions are not fully covered changes any results. From Figure~\ref{fig:galcover}, galaxies with $z \leq 0.02$ are excluded and the KS-testing described in Section~\ref{sec:kstesting} is run over this new sample of galaxies. This exclusion reduces the number of dwarf galaxies from 62,258 to 51,971, BPT galaxies from 387 to 326 and WHAN from 4,323 to 4,029.

The results do not differ from those in Table~\ref{tab:KStests} with the exception of BPT galaxies in 10NN. The p-value drops to 0.05 when comparing BPT galaxies versus non-AGN galaxies (\textit{'NOT'} galaxies). Raising the sampling size to 326 decreases the p-value even further suggesting that the distributions are \textit{not} similar. However, excluding nearby galaxies does not necessarily mean that it is an aperture effect. Only more energetic dwarf AGNs are visible at higher redshifts, so this may be a luminosity bias. Clearly, a more in-depth analysis is required to disentangle this result, but this is outside the scope of this study. This result does mean that it is not possible to conclusively rule out an environmental connection in dwarf BPT AGN galaxies.

For further analysis, low mass galaxies remain in the sample, but this presents another concern, namely whether the offset between the center of the galaxy and the fiber position is large enough for the fiber to not cover the core at all. The fiber will be fully offset if $\Delta\text{pos}>R_{\text{core}}+R_{\text{fiber}}$ which is the case for 8,103 of the 62,258 dwarf galaxies. Assuming the positions of the galaxies are equal to their core region, the spectra of these galaxies do not include their nuclei. However, most of these (8,100) are not linked to any spectroscopy runs and thus have no emission line fluxes. Of the 3 with emission lines, none of these galaxies appear in the BPT subsample but 2 of them appear in the WHAN subsample. Inspection of images of these WHAN galaxies with overplotted apertures reveals that their $R_{P90}$ are underestimated, but even then, the fiber position is not covering the core (by visual inspection). 
The NSA does list off-center SDSS spectroscopy as one of its caveats. However, only 2 out of all 4,323 WHAN galaxies are affected by this so this effect is ignored.

Regarding galaxies towards higher redshift, SF dilution (i.e weakening of the AGN signal due to an increasing ratio of emission by SF processes to AGN emission) increases since the fiber encloses an increasing fraction of the galaxies' total light. \citet{Moran2002} demonstrated that this effect biases against narrow line AGNs since that emission is drowned out by the host galaxy light. While no  AGN detection limits have been imposed in this study like the one used in \citet{Trump2015}, the importance of such methods appears crucial in studies focused on dwarf AGN selection improvement.

\subsection{On the environment and nearest neighbours}
\label{sec:envdiscussion}
Accepting the fact that environments do not affect AGN activity should not be done unconditionally as multiple studies have found connections (although sometimes weak) between AGN activity and environment  \citep{Miller2003, Kauffmann2004, Sabater2013, Amiri2019}. 
Some find that specific types of AGNs (e.g strong [O III] emitters or redder AGNs) are dependent on the local environment, so further subclassification of AGNs in dwarf galaxies may show a connection. However, classifying AGNs in the dwarf mass regime has a number of challenges. Emission line ratios and equivalent widths follow a mass trend (see Figure~\ref{fig:allAvgs}) and AGN characteristics in dwarf galaxies are hard to distinguish from e.g SF \citep{Trump2015}, which suggests that AGNs in the low mass-regime have to be treated differently. As noted by \citet{Dickey2019}, active dwarf galaxy samples  can include many false positives. This results in non-AGN galaxies being included in the AGN samples and the statistics would be biased towards regular galaxy distributions and thus not representative of the AGN population - an issue also raised by \citet{Hainline2016}

Furthermore, while the environment estimation  methods used in this study are tried and tested in other work for other purposes, there is a risk that they do not properly gauge the desired properties or the properties are not showing up in the statistics  due to obfuscating factors  such as trigger time lag \citep[the time delay it takes for AGN activity to begin after an interaction or harassment event, see e.g][]{Schawinski2007, Pimbblet2013, Shabala2012,Shabala2017} or SF contamination \citep{Trump2015}. Other environment estimation methods may reveal a connection to AGN activity, but the findings in this paper are in line with existing literature \citep[e.g][]{Miller2003, Kauffmann2004, Padilla2010, Sabater2013, Sabater2015, Man2019}. Therefore, using other relatively simple environment estimates will likely show similar results but more complicated ones such as the so-called tidal force estimator may find a difference in immediate environment \citep{Sabater2013}. Other options for improving the environment estimation method involves higher resolution and better spatially resolved observations  such as IFU surveys. They enable methods that more correctly gauge e.g recent merger history \citep[e.g][who inferred recent mergers from kinematically offset cores]{Penny2018}. 

Regarding galaxy tidal interactions, this study found no dependence in $\Delta$v to the nearest neighbour, although it is established that mergers can trigger AGN activity \citep[see e.g][]{Miller2003, Sabater2013, Ellison2019}. \citet{Treister2012} suggest that they are not necessary -- only for the brightest AGNs. As mentioned previously, using different methods such as distance to nearest bright neighbour \citep{Penny2016} or tidal force estimator \citep{Sabater2013}, this sample may show an excess of galaxy interactions. However, \citet{Sugata2019} found no excess of dwarf merger rate compared to regular merger rate for non-AGN dwarf galaxies suggesting that mergers are not important for AGN activity in the low-mass regime. 

\citet{Ellison2019} found that mergers can trigger AGN activity, though it may not be the dominant trigger. Furthermore, the fraction of disturbed galaxies are different depending on AGN selection method with mid-IR candidates being more often disturbed ($\sim 60\%$) than optical ones ($\sim 30\%$). This excess in mid-IR selected AGNs was also found by \citet{Satyapal2014}. Furthermore, \citet{Ellison2019} note that the excess of morphologically disturbed galaxies with AGN activity compared to disturbed non-AGN galaxies does increase with host mass and AGN luminosity. Conversely, the excess decreases towards lower mass galaxies giving credence to the notion that mergers are of lesser importance to AGNs in dwarf galaxies. 

However, the majority of if the galaxies in \citet{Ellison2019} are galaxies with $\text{log} M_*> 9.5$, and therefore extending the findings into the low mass regime should be done with care. The luminosity on O[ III] are also several orders of magnitude brighter. If the arguments from \citet{Ellison2019} are extended to dwarf galaxies, it is based on the assumption that AGNs in both low- and high mass galaxies are similar and can simply be scaled accordingly. 
Searching for mid-IR AGN dwarf galaxies has proven difficult as remarked by \citet{Lupi2020}. This means that a comparative study with \citet{Ellison2019} with dwarf galaxies is a difficult task.

Furthermore, the findings of \citet{Lupi2020} may point towards that AGNs in dwarf galaxies may be different from regular AGNs. It is remarked by e.g \citet{Mendez2013} and \citet{Azadi2017} that different wavelength diagnostics probe different AGN populations. It is therefore not possible to to conclude whether findings from \citet{Ellison2019} can be extended into the dwarf mass regime or not.

In this work, no restrictions were put on the neighbouring galaxy. Other work on this area such as \citet{Penny2016,Penny2018} required $M_k < -23$ of the neighbour since the mass of the neighbour decides how strong the tidal interactions are, and may be what is required to drive gas to the central region. Conversely,  a strongly disturbed dwarf galaxy may not have sufficient gas reservoirs to feed an AGN. From the method and results of this study alone, neither scenario is favoured. 

To explore this further, manual or automated visual inspection of the AGN sample may be required to give clues to properties such as morphology, nearby neighbours, and immediate environment. Morphology such as large-scale bars have found be correlated to AGN activity in e.g \citet{Galloway2015} (who suggested that a bar increases the probability of an actively accreting central black hole), while other studies such as \citet{Cheung2015} did not find this connection. While dwarf galaxies can be well-structured, they are often irregular (due to their low gravitational potential) and thus do not have a morphology that triggers AGN activity (e.g a bar). 

However, visual inspection of SDDS images of the active dwarf galaxies (specifically the majority of the \textit{'AND'} subsample, $N = 195$, see Section~\ref{sec:visinsp}) in this study has revealed no excess of morphology disturbances compared to a similar sized control sample (from \textit{'NOT}), which can be seen in Table~\ref{tab:visualInsp}. This is in line with what \citet{Sugata2019} found and \citet{Satyapal2014} noted that that optically selected AGNs do not tend to show an excess of mergers whereas mid-IR ones did. \citet{Goulding2018} found a similar excess in mid-IR data.

Complicating the morphology discourse further is the fact that \citet{Kruk2017} found that dwarf galaxies can be morphologically disturbed when found in isolation. What it means is that morphological disruptions of dwarf galaxies does not necessarily mean that they have been tidally affected or harassed by companion galaxies and as such, morphology is not an indicator of environment. 

Another important complication to consider is delayed triggering times of AGNs, which is suggested by e.g \citet{Pimbblet2013, Shabala2017}. The idea is that AGN activity does not start during an encounter or disturbance but rather 0.2-0.3 Gyr later. This timescale is the same order of magnitude as crossing time in rich galaxy clusters, which means that any present day AGN activity would be difficult to pin on a past event. \citet{Penny2018} also found dwarf galaxies with post-starburst spectra which supports the hypothesis that AGN activity can be delayed from an interaction since star bursts tend to be found in actively merging or harassed systems \citep{Hopkins2006}.

This would mean that the methods used in this study are not suitable to examine these properties. Other research such as \citet{Penny2018} found kinematically offset cores which could indicate accretion of IGM or merger, and the analysis of spatially resolved spectroscopy may be required to gauge the galaxy's past interaction history.  

\subsubsection{Weak galaxies}
\label{sec:weakDiscussion}
\begin{figure}
	\includegraphics[width=\columnwidth]{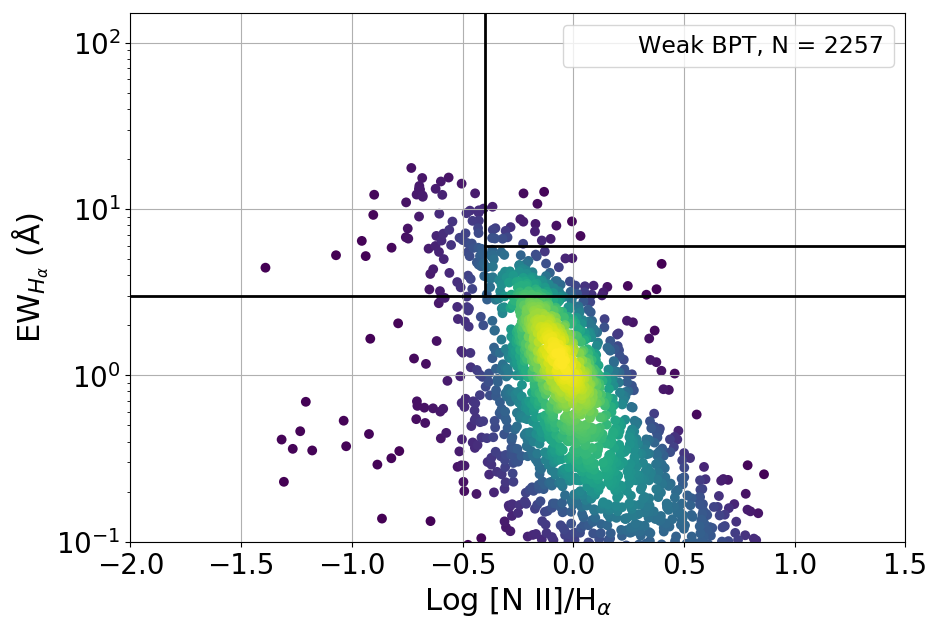}
	\includegraphics[width=\columnwidth]{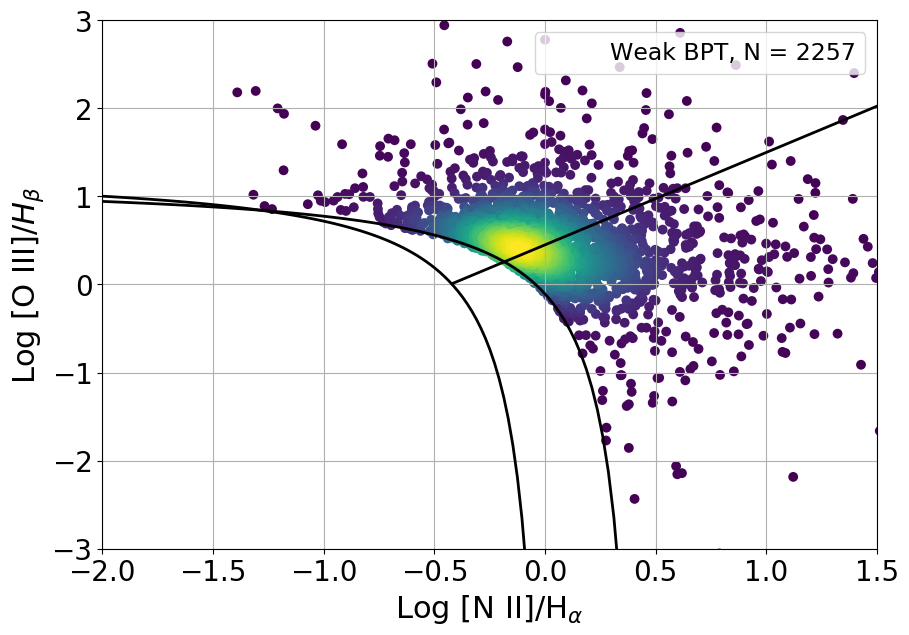}
	\caption{WHAN and BPT diagrams with weak BPT selected galaxies with dots colour-coded by their relative point density. These galaxies lie primarily in the retired region in the WHAN diagram while their positions in the BPT diagram are uncertain due to low S/N on the y-axis.}
	\label{fig:wBPTinWHAN}

\end{figure}
The only subsample to show a different environment is \textit{weak BPT galaxies}, although it is misleading to label them as BPT galaxies since their AGN activity is not definitive. As seen in Figure~\ref{fig:wBPTinWHAN}, the are primarily retired galaxies -- a classification defined in \citet{Stasinska2008}. \citet{Stasinska2008, Stasinska2015} discusses the implication that ionising radiation from evolved stellar population can place a galaxy in the LINER-region (and in the Seyfert-region, to a lesser extend) of the BPT diagram. They argue for using the equivalent width of H$_\alpha$ as a method to break the degeneracy of the BPT classified LINERs and retired galaxies. 

Although most of the weak BPT galaxies are in the BPT-Seyfert region (Figure~\ref{fig:wBPTinWHAN}), their exact positions are uncertain as their S/N is too low for proper line ratio measurements. Their horizontal positions, though, are mostly correct as measurements of H$_\alpha$ and N[II] are of good quality. This would put most of them in the composite or LINER region of the BPT diagram.

It is not surprising to find retired galaxies in denser environments as the connection between dense environments and SF quenching is well-established \citep{Balogh1998, Lewis2002, Peng2010, Peng2012, Penny2016, Spindler2018}. A number of processes such as ram-pressure stripping, heating of cold gas reservoirs, and galaxy harassment have been found theorised to quench star formation and these processes are more likely to happen in dense environments. Therefore, this subsample behaves as one would expect.

\subsection{On selection method bias}
\label{sec:discSelectionMethod}
A concern in this study is the extension of regular optical AGN diagnostic in to dwarf mass regime -- a concern explored in detail in e.g \citet{Cann2019} -- and studies like \citet{Trump2015},  \citet{Penny2018}, and \citet{Dickey2019} does bring into question whether AGNs in dwarf galaxies are robustly identified in regular diagnostic tools.

One observation from this study is that there a clear mass trends in both BPT and WHAN, and extending these diagnostic into the low-mass regime may carry biases, which are often not corrected for. As shown in e.g \citet{Reines2013, Sartori2015, Stasinska2015, Baldassare2018, Dickey2019, Cann2019}, a number of AGNs will not be identified with standard optical AGN diagrams even though they clearly show AGN characteristics in other diagnostics (e.g X-ray or mid-IR selecttion), or numerous non-AGN will be included in a sample if the selection criteria are too lenient -- something that is exaggerated in low-mass galaxies \citep[][]{Stasinska2015, Trump2015, Hainline2016}.  
 
\citet{Trump2015} suggest that SF dilution biases against AGN in low mass galaxies. The emission lines are drowned out by SF radiation and also suggest that AGNs are fueled by the same gas as SF resulting in SF 'stealing' available gas from the AGN and preventing it from reaching very high energy outputs. The consequence is that low mass galaxies have weaker relative AGN emission compared to high mass galaxies. In order to correct for this, one solution could be to mass-weigh emission line ratios (and equivalent widths). However, it is well-established that AGN fraction is strongly correlated with host mass, and finding a correction factor on this parameter may prove difficult.
However, this mass bias may be caused by selection biases due to aperture (see Section~\ref{sec:fiberbias}) -- something that \citet{Moran2002} argue.

Studies in other wavelengths may reveal further AGNs in dwarf galaxies, and this has been explored in e.g \citet{Lupi2020,Birchall2020} in mid-infrared and x-ray respectively. However, \citet{Lupi2020} remark that the poor resolution of mid-infrared surveys and contamination makes this wavelength regime a bad choice for identifying AGNs in dwarfs. \citet{Birchall2020} found that out of 4,331 dwarf galaxies, 61  show AGN activity, and 85\% of these identified AGNs did not show up in optical wavelengths. This suggests that x-ray data is suitable to complement optical data in search of dwarf AGNs. Unfortunately, the x-ray data and coverage of the sky is limited, but it may be used in conjunction with optical data sets to find a potential correction function for optical diagnostics.

\citet{Baldassare2018} used long-term optical variability to identify dwarf AGN and noted that star formation dilution and low metallicity may be likely reasons why AGNs are missed in dwarf galaxies. Using nuclear variability to identify AGNs would circumvent problems with correlating observations across wavelength regimes or mass-weighing emission lines --two methods which carry limitations discussed earlier in this section. The obvious downside to this approach is the requirement of observations at different times over several years \citep[data from][spans over ~5 years]{Baldassare2018} and the need to spatially resolve cores of dwarf galaxies, which are very small.

Another finding in this study is that the WHAN diagram tends to classify low mass galaxies with low S/N on H$_\beta$ and/or [O III] as retired (Figure~\ref{fig:wBPTinWHAN}).  This suggests that using the WHAN diagram on BPT selected AGNs is a fast way of identifying contaminating retired galaxies in AGN samples -- at least in low-mass samples. This is assuming that WHAN classified AGNs are indeed AGNs. The AGN fraction as function of stellar mass (Figure~\ref{fig:AGNFraction}) shows that the two methods find different AGN fractions and follow different mass trends. As noted by \cite{Cid2010}, WHAN tends to probe weaker AGNs, which obviously are more common in lower stellar mass galaxies. 

\section{Conclusions}
\label{sec:conclusions}
The main discussion points will be summarised in this section. For a short summary, skip to the end. This study finds that the environments of AGN dwarf galaxies are no different than the environments of regular dwarf galaxies regardless of AGN selection method. There is neither a difference in local galactic density nor a velocity difference to the nearest neighbours suggesting that the main AGN trigger is an internal process. However, the non-dependence found can be a result of the method -- either from biases in  the AGN selection methods or from not taking various factors such as time delay, mass trends, or SDSS fiber aperture bias into account.

For example, using only galaxies with $z \geq 0.02$ -- basically galaxies whose whole core region is covered by the SDSS fiber -- the environments of BPT galaxies are distinguishable enough from a matched control sample of non-AGN galaxies to show up as statistically significant. However, this effect may be due to other reasons than just fiber aperture such as e.g luminosity. Without the redshift restriction, the only subsample to show a difference in environment is weak BPT galaxies that show up as retired galaxies in WHAN. This subsample prefers denser environments, which makes sense because this subsample most likely consists of quenched star formation and dense environments are known to cause star formation quenching.

The analysis also looked at other galactic parameters. The distributions of stellar mass, redshift, and r-magnitudes were used as a test to see how they compare to existing literature. The stellar masses and magnitudes behave as expected with AGNs tending to be brighter and more frequent in higher mass galaxies. The redshift distributions between the samples are slightly different with WHAN tending to be found at higher redshifts and BPT at lower redshifts compared to the regular galaxies. This might be due to observational effects -- BPT requires high quality measurements of weak lines and thus favours brighter and closer galaxies while WHAN probes weaker AGNs that are harder to detect at higher redshift.

The environment description methods only probe the galaxies in their current environment, though, and does not take their past into account. Some research suggests a time delay of $\sim 0.2-0.3$ Gyr \citep{Pimbblet2013, Shabala2017}. Strong encounters in the past might have left an impression on the galaxies' morphologies, but visual inspection did not reveal any significant disturbances. This is in line with e.g \citet{Sugata2019}, who did not find an excess of disturbed or merging dwarf AGN galaxies compared to a control sample. Other indicators might exist of past encounters or significant disturbances such as kinematically offset cores, so other diagnostics may be needed for better analysis and understandings.

As an attempt to avoid bias from the selection method, two AGN selection methods were used. However, the samples from both methods were indiscernible from each other and from regular dwarf galaxies regarding environmental analysis, which means that optical AGN features are not affected by the environment. The two methods themselves seem to probe slightly different galaxy populations. While most BPT galaxies are also identified as AGNs in WHAN ($195/296 ~ 66\%$), the majority of WHAN galaxies are classified as star-forming galaxies or composite ones in BPT ($4099/4294 ~ 95\%$). However, this does not mean the WHAN AGN classification is untrustworthy. The advantage of WHAN is that it aims to probe weaker AGNs and as several studies have found, AGNs in dwarf galaxies can be diluted by star formation \citep{Trump2015}, be rejected because of intrinsic weak emission lines \citep{Cid2010}, or have higher sensitivity to environment \citep{Wetzel2013} -- effects that all may weaken AGN signatures. 

Regardless of AGN selection method, neither local nor immediate environment seem to play a role in triggering AGNs in dwarf galaxies judging from the similarity of environment between AGNs and non-AGNs. While this is in agreement with existing literature, there are a number of factors that weakens this conclusion. Firstly, whether the environment estimates actually gauge the desired properties can be called into question. It could also be that the observable environment parameters have changed since they first triggered the AGN activity. Secondly, the sample of AGNs are found using diagnostic tools developed for regular galaxies, and extending these to the low-mass regime carries biases that results in an unpure sample consisting of many non-AGNs. 

A solution to the first issue would be to develop more complicated environment description tools as already seen in e.g \citet{Sabater2013}, \citet{Baldassare2018}, and \citet{Penny2018}. Generally, these methods can be thought of as having a longer lookback time compared to simpler methods. Regarding the second issue, other wavelength regimes can help identify a large fraction of optically undiscovered AGNs \citep[in][, optical diagrams failed to find 85\% of AGNs]{Birchall2020} or variability surveys \citep[e.g][]{Baldassare2018} also help. A third option is to mass-weigh emission line ratios -- an option motivated by the fact that there are mass trends in optical diagnostic diagrams.

These findings can be summarised as follows:
\begin{itemize}
    \item There is no difference in neither local or immediate environment between AGN dwarf galaxies and non-AGN dwarf galaxies suggesting that the environment does not play a role in triggering AGN activity.
    \item This non-dependence on environment was found regardless of selection method (BPT and WHAN), although a redshift-limited (due to SDSS fiber coverage) sample of BPT galaxies did show a difference in environment. Thus it is not possible to conclusively rule out an environmental dependance of BPT galaxies.
    \item Concerns were raised regarding both the AGN selection methods and the environment selection methods. AGN diagnostics are calibrated to regular galaxies, and extending these into the low-mass galaxy regime may produce samples either including many non-AGN galaxies or excluding AGNs.
    \item Regarding the environment, the utilised methods probe the current environment, but the current environments of galaxies may not be the environments that triggered AGN activity -- there may be a time delay before the onset of AGN activity
    \item For future work on the subject of AGNs in dwarf galaxies, involves calibration of AGN diagnostics such as the BPT and WHAN diagrams to low-mass galaxies. This may include using other wavelength regimes in the identification of dwarf AGN or develop a mass-weighing factor on emission line ratios since they trend with stellar mass.
\end{itemize}

\section*{Acknowledgements}
We appreciate the thorough and insightful comments by the reviewers that helped improve the paper, especially in regards to aperture bias and constructing control samples.
Furthermore, KAP acknowledges the support of the Science and Technology Facilities Council (STFC) through the University of Hull’s Consolidated Grant ST/R000840/1. MTK acknowledges the support of University of Hull Astrophysical Data Science Cluster.

Funding for SDSS-III has been provided by the Alfred P. Sloan Foundation, the Participating Institutions, the National Science Foundation, and the U.S. Department of Energy. The SDSS-III web site is http://www.sdss3.org.
SDSS-III is managed by the Astrophysical Research Consortium for the Participating Institutions of the SDSS-III Collaboration including the University of Arizona, the Brazilian Participation Group, Brookhaven National Laboratory, University of Cambridge, University of Florida, the French Participation Group, the German Participation Group, the Instituto de Astrofisica de Canarias, the Michigan State/Notre Dame/JINA Participation Group, Johns Hopkins University, Lawrence Berkeley National Laboratory, Max Planck Institute for Astrophysics, New Mexico State University, New York University, Ohio State University, Pennsylvania State University, University of Portsmouth, Princeton University, the Spanish Participation Group, University of Tokyo, University of Utah, Vanderbilt University, University of Virginia, University of Washington, and Yale University.

\section*{Data availability}
The data underlying this article is the NASA-Sloan Atlas (\texttt{v0\_1\_2}) and can be accessed at \url{nsatlas.org}. 




\bibliographystyle{mnras}
\bibliography{bibliography} 

\bsp	
\label{lastpage}
\end{document}